\documentstyle[aps,prl,twocolumn,graphicx,amsmath]{revtex}

\begin{document}
\newcommand{\J}{{\mathbf J}}
\newcommand{\du}{{\mathbf w}}
\newcommand{\T}{{\mathbf T}}
\newcommand{\B}{{\mathbf B}}
\newcommand{\s}{{\mathbf S}}
\newcommand{\ksi}{{\boldsymbol \xi}^{\mu}}
\newcommand{\bsxi}{\boldsymbol \xi}
\newcommand{\qh}{\hat{q}}
\newcommand{\Qh}{\hat{Q}}
\newcommand{\sgn}{\text{sgn}}
\newcommand{\eps}{\varepsilon}
\newcommand{\al}{\alpha}
\newcommand{\lan}{\langle\langle}
\newcommand{\ran}{\rangle\rangle}
\newcommand{\btau}{{\boldsymbol \tau}}

\def\lim{\mathop{\rm lim}}
\def\extr{\mathop{\rm extr}}
\def\Tr{\mathop{\rm Tr}}

\twocolumn[\hsize\textwidth\columnwidth\hsize\csname@twocolumnfalse\endcsname
\title
{Multilayer neural networks with extensively many hidden units}
\author
{Michal Rosen--Zvi$^1$, Andreas Engel$^2$, and Ido Kanter$^1$ }
\noindent
\address{$^1$ Minerva Center and Department of Physics, Bar-Ilan University, 
   Ramat-Gan, 52900 Israel\\
 $^2$ Institut f\"ur Theoretische Physik, Otto-von-Guericke Universit\"at, \\
         PSF 4120, 39016 Magdeburg, Germany}
 
\maketitle

\begin{abstract}
The information processing abilities of a multilayer neural network with a
number of hidden units scaling as the input dimension are studied using 
statistical mechanics methods. The mapping from the input
layer to the hidden units is performed by general symmetric Boolean functions
whereas the hidden layer is connected to the output by either discrete or
continuous couplings. Introducing an overlap in the space of Boolean 
functions as order parameter the storage capacity if found to scale with the
logarithm of the number of implementable Boolean functions. The generalization
behaviour is smooth for continuous couplings and shows a discontinuous 
transition to perfect generalization for discrete ones. 
\end{abstract}
\pacs{} ]


Statistical mechanics investigations of artificial neural networks continue to
play a stimulating and integrating role in the scientific dialogue between
discipline as diverse as neurophysiology, mathematical statistics, computer
science and information theory. In particular the study of feed-forward neural
networks pioneered by Gardner \cite{Gardner} has revealed a
variety of interesting results on how these system may learn different tasks
of information processing from examples (for a review see \cite{EnvB}). Of
particular importance in this respect are multilayer networks (MLN) because of
their ability to implement any function between input and output \cite{Cyb}
which makes them attractive candidates for many practical
applications. It is well known that very many hidden 
units are needed in order to realize this vast computational
complexity. However, statistical mechanics studies of MLN have so far been
mostly restricted to systems with very few hidden units as compared to the
number of inputs \cite{smMLN}. In the present letter we overcome this
limitation and study the storage and generalization abilities of a
tree MLN in which the size of the hidden layer scales in the same way as the
input dimension.  


We consider a MLN with $N$ binary hidden units $\tau_i=\pm 1, i=1,...,N$ 
feeding a binary output $\sigma=\sgn(\sum_i J_i \tau_i)$ through a coupling 
vector $\J=J_1,...,J_N$. The hidden units are determined via Boolean functions
$\tau_i=B_i(\s_i)$ by disjoint sets of inputs $\s_i=S_{i1},...,S_{iL}$
containing $L$ elements each. We are interested in the limit $N\to\infty$ with
$L$ remaining constant. 

In order to keep the connection with neural network architectures we restrict
ourselves to {\it symmetric} Boolean functions characterized by
$B_i(-\s_i)=-B_i(\s_i)$. There are $2^{2^{L-1}}$ 
such functions with $L$ inputs, with only few of them realizable by a coupling
vector $\du_i$ according to $B_i(\s_i)=\sgn(\sum_j w_{ij} S_{ij})$. For $L=3$
there are, e.g., 16 symmetric Boolean functions but only 14 of them are
linearly separable.


In order to investigate the storage and generalization properties of the
network we consider a set of $\al L N$ inputs $\ksi_i,\mu=1,...,\al LN$ the 
components $\xi_{i1}^{\mu},...,\xi_{iL}^{\mu}$ of which are independent,
identically distributed 
random variables with zero mean and unit variance. We then ask for the ability
of the network to map these inputs on outputs $\sigma^{\mu}=1$ for all $\mu$ by
adapting the Boolean functions $B_i$ and the couplings $J_i$ appropriately. 

The central quantity in the statistical mechanics analysis is the
{\it quenched entropy}
\begin{equation}\label{defs}
  s=\lim_{N\to\infty}\frac{1}{N} \lan \int d\mu(\J) \Tr_{\{B_i\}}
    \prod_{\mu=1}^{\al L N}\theta(\sum_i J_i B_i(\ksi_i))
       \ran_{\{\ksi_i\}}
\end{equation}
where $d\mu(\J)$ is the proper measure in the space of couplings $\J$, the
trace denotes the sum over all Boolean functions, the product 
is non-zero only if the arguments of all of the $\theta$-functions is positive
and the double angle stands for the average over the inputs. The
determination of $s$ can be performed using the replica trick and introducing
the overlap between two solutions in the combined space of couplings $\J$ and
Boolean functions $B_i$ of the form 
\begin{equation}\label{defq}
q^{ab}=\frac{1}{N}\sum_i J_i^a J_i^b \;\lan B_i^a({\bsxi})
        B_i^b({\bsxi})\ran_{\bsxi}
\end{equation}
with the average being now over a single, $L$-component vector
$\bsxi$.  
Exploiting the fact that this average involves a finite number of terms only
and assuming replica symmetry $q^{ab}=q$ for $a\neq b$ we can write $s$ using
standard techniques \cite{EnvB} in the form  
\begin{equation}\label{ress}
  s=\extr_{q, \qh}\left[ G_C(q, \qh)+G_S(\qh)+\al L G_E(q)\right],
\end{equation}
with the explicit expressions for the functions $G_C, G_S$ and $G_E$ depending
on $L$, the constraints on $\J$ and on whether the storage or the
generalization problem is addressed.


Let us begin with the storage problem by asking for the storage capacity
$\al_c$ defined as the maximal value of $\al$ for which the system can still
realize all desired input-output mappings with probability 1. Performing the 
replica limit with the number of replicas tending to zero characteristic for
this problem we find 
\begin{equation}\label{GEstor}
  G_E(q)=\int Dt \ln H(Q\;t)
\end{equation}
with the abbreviations $Dt=dt\, e^{-t^2/2}/\sqrt{2\pi}$, 
$H(x)=\int_{x}^{\infty} Dt$, and $Q=\sqrt{q/(1-q)}$. The expressions for $G_C$
and $G_S$ depend on the constraints on the coupling vector $\J$. 


A particular simple case is given by Ising couplings $J_i=\pm 1$. From the
symmetry of the Boolean functions considered it is clear that it
is sufficient to consider $J_i=1$ for all $i$. Consequently in this case all
flexibility of the network rests in the 
choice of the Boolean functions between input and hidden layer and 
$q$ is a sole overlap in the space of these Booleans. We find 
$G_C=\qh(1-q)/2$ where $\qh$ denotes the conjugate order parameter to
$q$. Moreover, in the case where all $2^{2^{L-1}}$ symmetric  
Boolean functions are admissible we use the identity
\begin{equation}
  \Tr_{\{B_i\}}\exp(\sqrt{\frac{\qh}{2^{L-1}}} \nonumber
       \sum_{\mathbf\xi} z_{\mathbf\xi} B_i({\mathbf\xi}))
  =\prod_{\mathbf\xi}(2\cosh(\sqrt{\frac{\qh}{2^{L-1}}}\; z_{\mathbf\xi}))
\end{equation}
with the sums and products over ${\bsxi}$ running  over all $2^{L-1}$
configurations of $\bsxi$ with $\xi_1=1$ to find
\begin{equation}
  G_S=2^{L-1}\int Dz \ln \left[ 2\cosh(\sqrt{\frac{\qh}{2^{L-1}}}\;z)\right].
\end{equation}
Under the transformations $\qh\mapsto 2^{L-1}\qh$ and $\al\mapsto
2^{L-1}\al/L$ the resulting expression for the entropy maps {\it exactly} on
the result for the Ising perceptron corresponding to $L=1$ and we may 
therefore
use the well known results for this case \cite{KrMe}. Accordingly the storage
capacity is overestimated by the replica symmetric expression and the correct
result   
\begin{equation}\label{h1}
\al_c(L)= \al_c(1)\;2^{L-1}/L \cong 0.83\; 2^{L-1}/L
\end{equation}
is given by the value of $\al$ at which the entropy $s(\al)$ turns negative. 
The storage capacity is hence proportional to the {\it logarithm} of the 
number of implementable Boolean functions. This result is in
accordance also with the rigorous upper bound $\al_c\leq2^{L-1}/L$ resulting
from the annealed entropy $s^{\text{ann}}=(2^{L-1}-\al L)\ln 2$. 
As in the case of
the Ising perceptron this bound is related to information theory.
The full specification of the network with all $J_i=1$ requires 
$N\,2^{L-1}$ bits of information necessary to pin down the $N$ Boolean 
functions $B_i$. Therefore the machine cannot store more than $N\,2^{L-1}$
bits and $\al_c$ cannot exceed $2^{L-1}/L$.

\begin{figure}[tb]\vspace*{-.75cm}
\hspace*{-1cm}\includegraphics[width=8cm,angle=270]{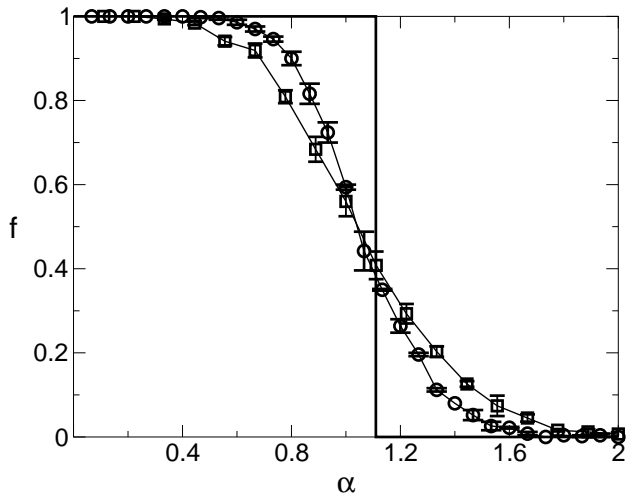} 
   \caption{\label{Isingcap} Fraction $f$ of $3 \al N$ random input-output
          mappings implementable by 
  a MLN with $3N$ inputs and $N$ hidden units as function of $\al$ for $N=3$
  (squares) and $N=5$ (circles). The couplings between hidden units and output
  are fixed to $J_i=1$ for all $i$ and enumerations are performed over all 
  combinations of symmetric Boolean functions $B_i$ between input and hidden
  layer. For every value of $\al$, 200 realizations of Gaussian inputs
  where averaged over. The solid line gives the analytical result describing
  the limit $N\to\infty$.} 
\end{figure}

Fig.\ref{Isingcap} compares the analytical result $\al_c(3)\cong 1.11$ for
$L=3$ with numerical simulations using exact enumerations. Even for the small 
sizes accessible to this numerical technique we find a steepening of the 
transition with increasing $N$ and a crossing point of the curves near to the 
theoretical prediction. 

If the trace over the Boolean functions in (\ref{defs}) is restricted to those
which can be realized by perceptrons with coupling vectors $\du_i$ the exact
mapping on the Ising perceptron no longer holds. Solving the
corresponding extremum conditions numerically for $L=3$ we find $\alpha_c
\cong 1.06$ for this case. The reduction of $\al_c$ compared to the
unrestricted case is roughly as the reduction in the logarithm of the number
of admissible Boolean functions $1.06/1.11\cong\ln(14)/\ln(16)$.

\begin{figure}[tb]\vspace*{-.75cm}
\hspace*{-1cm}\includegraphics[width=8cm,angle=270]{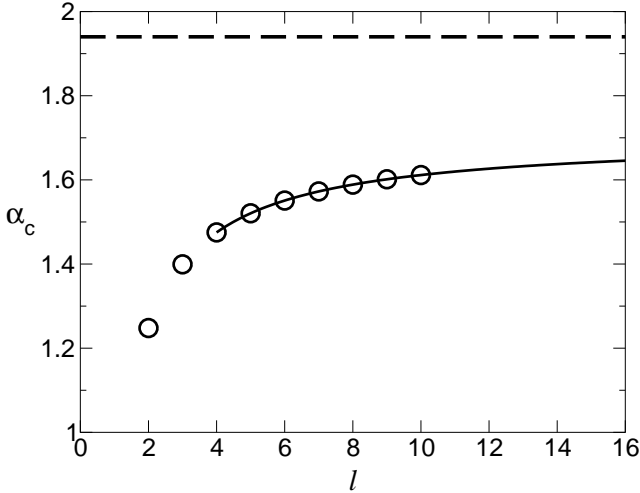} 
   \caption{\label{CDigital} Storage capacity of a MLN with $N\to\infty$
     hidden units and $3N$ inputs with couplings $J_i$ between hidden layer
     and output taking $2l$ discrete values. The inputs are mapped to the
     hidden layer by symmetric Boolean functions $B_i$. The solid line is the 
     fit $\al_c\sim 1.70-0.91/l$ 
     to the asymptotic behavior, the dashed line gives the replica
     symmetric result for continuous couplings $J_i$.}
\end{figure}

It is possible to generalize the above analysis to the case of discrete
couplings with finite synaptic depth $l$ of the form 
$J_i=\pm 1/l,\pm 2/l,..., \pm 1$ by building on the analysis of the analogous
case for the perceptron \cite{GuSt,Kanter}. In this case the additional order
parameter $\bar{q}=\sum_{i}(J_i^a)^2/N$, and its conjugate, 
$\hat{\bar{q}}$ have to be introduced. For $G_E$ we then again find
(\ref{GEstor}) with now $Q=\sqrt{q/(\bar{q}-q)}$. Moreover 
$G_C=-\hat{\bar{q}}\bar{q} + \qh q/2$ and, if all symmetrical
Boolean functions are admissible,
\begin{align}
G_S&=\int\prod_{\bsxi} Dz_{\bsxi} \nonumber\\
   & \ln \Tr_J \exp(-(\frac{\qh}{2}-\hat{\bar{q}})J^2) \nonumber
   \prod_{\mathbf\xi} 2\cosh(J\,\sqrt{\frac{\qh}{2^{L-1}}}\; z_{\mathbf\xi}),
\end{align}
with $\Tr_J$ denoting the trace over the $2l$ possible values of the couplings
$J_i$. 
Using these results we have numerically calculated the storage capacity
$\al_c(l)$ for the simplest case $L=3$ as a function of the synaptic depth
$l$. The results are shown in fig.\ref{CDigital} together with a fit to the
asymptotic behavior. The capacity increases from $\alpha_c\cong 1.11$ of 
the Ising case, $\l=1$, to roughly $1.7$ for large $l$. It is rather difficult
to compare these analytical findings with numerical simulations since the
effects of the finite synaptic depth do not show up at the small values of
$N$ accessible to exact enumerations \cite{PBGDK}. 

To complete the analysis of the storage properties 
we analyze the case of continuous couplings $\J$  between hidden
and output layer. It is convenient to eliminate the additional order parameter
$k$ necessary in this case to enforce the normalization $\J^2=N$ by
introducing $\Qh=\qh/(k+\qh)$. Within replica symmetry the quenched entropy
$s$ is then again of the form (\ref{ress}) with $G_C=0$, $G_E$ given by 
(\ref{GEstor}), and the extremum taken now over $Q$ and $\Qh$. Moreover 
\begin{align}
  G_S=&\frac{1}{2}\ln(1-\frac{\Qh}{1+Q^2})\nonumber\\
      &+\int\prod_ {\bsxi} Dz_{\bsxi}\ln \Tr_B\;\exp(\frac{\Qh}{2^L}\;
       (\sum_{\bsxi} z_{\bsxi} B({\bsxi}))^2).
\end{align}
The storage capacity $\al_c$ can be obtained from these expressions in the
limits $Q\to\infty, \Qh\to\infty$ corresponding to $q\to 1$. This limit 
indicates that different solutions of the storage problem may at most 
differ in a non-extensive number of components $J_i$ and Boolean functions 
$B_i$. We then find $G_E\sim -Q^2/4$ and, if all
Boolean functions are admissible, $G_S\sim \Qh(1/2+(2^{L-1}-1)/\pi)$ giving
rise to 
\begin{equation}\label{alspherrs}
  \al_c^{RS}=\frac{2+\frac{4}{\pi}(2^{L-1}-1)}{L}.
\end{equation}
For $L=3$ this yields $\al_c^{RS}=2/3+4/\pi\cong 1.94$. If only linearly
separable Boolean functions implementable by coupling vectors $\du_i$ are
considered the asymptotic behavior of $G_S$ is more difficult to obtain. For
the  case $L=3$ we find $\al_c\cong 1.85$. Again the relative reduction of
$\al_c$ when compared to the unrestricted case is roughly given by the ratio
of the logarithms of the number of available Boolean functions per hidden
unit.  

It is possible to derive an upper bound for $\al_c$ as has been done for MLN
with a finite number of hidden units \cite{MiDu} by using some exact
results for the perceptron \cite{Cover}. For $L=3$ we find 
$\al_c(L=3)\leq 2.394$ and the replica symmetric result is therefore within 
the bound. For large $L$ the bound is given by 
$\al_c(L\to\infty)\lesssim 2^{L-1}/L+\ln 2$ and shows the same scaling with 
$L$ as (\ref{alspherrs}).
Nevertheless the replica symmetric result (\ref{alspherrs}) is very likely 
to overestimate the storage capacity as can be seen from fig.\ref{CDigital} in
which the result for $\al_c$ for $L=3$ is included as horizontal dashed line. 
Unlike the case of the perceptron \cite{GuSt} the values for $\alpha_c$ 
for finite synaptic depth seem not to approach the value for continuous 
couplings when $l\to\infty$. It would hence be very interesting to 
investigate the implications of replica symmetry breaking, both on the case 
of continuous couplings and of couplings with finite synaptic depth 
\cite{Robert}.


Let us finally elucidate the generalization problem, i.e. the ability of
the network to infer a rule from examples. To this end we consider as usual
two networks of the same type with the couplings and Boolean function of one
of them (the ``teacher'') fixed at random. The other network (the ``student'')
receives a set of randomly chosen inputs $\ksi_i,\mu=1,...,\al L N$ together
with the corresponding outputs $\sigma^{\mu}_T$ generated by the teacher. The
task for the student is to imitate the teacher as well as possible. The 
success in doing so is quantified by the generalization error $\eps$ defined 
as the probability that a
{\it newly} chosen random input is classified differently by teacher and
student. 

As is well known the statistical mechanics analysis of the generalization
problem builds again on the expression (\ref{defs}) for the quenched entropy 
with the number of replicas now tending to 1 rather than to 0 
\cite{OpHa,EnvB}. A nice feature of this limit is that replica symmetry is 
known to be stable. The order parameter $q$ defined in
(\ref{defq}) now gives the typical overlap between teacher and student and 
determines the generalization error $\eps$ in a simple way. In the present
situation we have the standard relation $\eps=(\arccos q)/\pi$. Moreover
(\ref{GEstor}) is replaced by 
\begin{equation}\label{GEgen}
    G_E(q)=2 \int Dt\; H(Q\;t)\;\ln H(Q\;t).
\end{equation}
The case using Ising couplings $J_i=\pm 1$ and all symmetric Boolean functions
can again be mapped exactly on the Ising perceptron. Correspondingly there is
a {\it discontinuous} transition to perfect learning, $\eps=0$ for $\al>\al_d$
\cite{Gyo} with $\al_d=1.24\;2^{L-1}/L$. This transition occurs when all
Boolean functions of the student ``lock'' onto the corresponding input-hidden
mappings of the teacher and is also expected to occur in the case where only 
a restricted set of Booleans can be implemented. 

For continuous couplings we find $G_C=Q^2\Qh/((1+Q^2)(1-\Qh))$ and 
\begin{equation}\label{GSgen}
  G_S=\frac{1}{2}\ln(1-\Qh)+\sqrt{1-\Qh}
      \int\prod_ {\bsxi} Dz_{\bsxi}\; g(z_{\bsxi})\,\ln g(z_{\bsxi})
\end{equation}
where
\begin{equation}\label{GSh}
 g(z_{\bsxi})=\Tr_B \exp(\frac{\Qh}{2^L}(\sum_{\bsxi}\,z_{\bsxi} B(\bsxi))^2).
\end{equation}
For small $\al$ this gives rise to $\eps\sim 1/2-\al L/(\pi^2 2^{L-1})$ which 
coincides with the result for the perceptron for L=1 as it should. With
increasing $L$ the initial decay of the generalization error becomes slower
reflecting the increasing complexity and storage abilities of the network. 
There is no retarded learning because of the non-zero 
correlation between the hidden units and the output \cite{retlearn}.  
For large $\al$ the generalization behaviour is dominated by the fine tuning 
of the student couplings between hidden layer and output to the respective 
couplings of the teacher resulting in the ubiquitous power law decay 
$\eps\sim 0.625/(L\al)$.

In conclusion we have quantitatively characterized the storage and 
generalization abilities of a multilayer neural network with a number of 
hidden units scaling as the input dimension. If the mapping from the input 
to the hidden layer is realized by symmetric Boolean functions with $L$
inputs the capacity is found to be proportional to the logarithm of the 
number of these Boolean functions divided by $L$. The more conventional case 
in which the hidden units are the outputs of perceptrons with couplings 
$\du_i$ is more difficult to analyze. However, speculating that the above 
scaling holds true also in this case and observing that the logarithm of 
the number of Boolean functions which can be implemented by a perceptron with 
$L$ inputs is $O(L^2)$ we arrive at the interesting result that the number 
of stored input-output relations {\it per weight} of the network is 
proportional to $L$. This implies that doubling the number of couplings 
in the network would increase the storage capacity by a factor of 2 making 
the proposed architecture superior to MLN with few ($K\ll N$) hidden 
units in which the storage capacity is known to increase at most 
logarithmically with the number of weights.

\vspace*{.5cm}

{\bf Acknowledgment:} We have benefitted from discussions with Wolfgang 
Kinzel, Robert Urbanczik, Peter Reimann, and Stephan Mertens. 
We would like to thank the 
Max-Planck-Institut f\"ur Physik komplexer Systeme in Dresden where this 
work was finished for hospitality and the GIF for support.

\end{document}